\documentclass[10pt]{iopart}
\usepackage{epsf,citesort}

\newcommand{\gl}[1]{Eq.~(\ref{#1})}
\newcommand{\gls}[2]{Eqs.~(\ref{#1},\ref{#2})}

\def\gtrless{\raise2.5pt\hbox{$>$}\llap{\lower2.5pt\hbox{$<$}}}
\def\gtrapprox{\raise2.5pt\hbox{$>$}\llap{\lower2.5pt\hbox{$\approx$}}}
\def\gtrapprox{\raise2.5pt\hbox{$>$}\llap{\lower2.5pt\hbox{$\approx$}}}
\def\lessapprox{\raise2.5pt\hbox{$<$}\llap{\lower2.5pt\hbox{$\approx$}}}

\newcommand{\beq}[1]{\begin{equation}\label{#1}}
\newcommand{\eeq}{\end{equation}} 
\newcommand{\for}{\qquad\mbox{for }\quad}

\newcommand{\with}{\qquad\mbox{with }\quad}

\renewcommand{\rho}{\varrho}

\renewcommand{\vec}[1]{\, \mathbf{#1}}
\newcommand{\uvec}[1]{\,\hat{ \mathbf{#1}}}
\newcommand{\avg}[1]{\langle #1 \rangle}
\newcommand{\gva}[3]{ #1\rangle #2 \langle #3}
\newcommand{\ket}[1]{ #1 \rangle}

\newcommand{\scal}[2]{\vec {#1}\cdot \!\vec {#2} }

\begin{document}

\title[Hydrodynamic boundary conditions]{
Statistical mechanics derivation of hydrodynamic boundary
conditions: the diffusion equation\footnote[4]{Dedicated to
Jean Pierre Hansen on the ocassion of his 60th birthday}}

\author{ M. Fuchs\ddag
\footnote[3]{
Permanent address: Physik-Department, Technische Universit\"at
M\"unchen, James-Franck-Str., 85747 Garching, Germany} and K. Kroy\dag\ }

\address{\dag\ Department of Physics and Astronomy, The University of
Edinburgh, JCMB King's Buildings, Edinburgh EH9 3JZ, GB}

\address{\ddag\ Institut Charles Sadron, 6, rue Boussingault, 67083
Strasbourg Cedex, France}

\begin{abstract}
Considering the example of interacting Brownian particles we present a
linear response derivation of the boundary condition for the
corresponding hydrodynamic description (the diffusion equation).  This
requires us to identify a non-analytic structure in a microscopic
relaxation kernel connected to the frequency dependent penetration
length familiar for diffusive processes, and leads to a microscopic
definition of the position where the hydrodynamic boundary condition
has to be applied. Corrections to the hydrodynamic limit are obtained
and we derive general amplitudes of spatially and temporally long
ranged fluctuations in the considered diffusive system.
\end{abstract}

\maketitle

\section{Introduction}

The description of dynamical processes in condensed matter greatly
simplifies if fluctuations are studied which are slow and smooth
compared to the microscopic length and time scales of the system. Then
hydrodynamic equations for a small number of fields can be derived,
either using rather general phenomenological considerations, or by
coarse graining starting from a microscopic statistical mechanics
description. In the latter a large number $N$ of particles needs to be
handled and the hydrodynamic fields (normally) arise as coarse grained
densities of conserved variables \cite{hansen}.  The Zwanzig--Mori
operator formalism enables one to perform the coarse graining of the
microscopic equations of motion using spatially Fourier transformed
variables in the limit of vanishing wavevector $q$ (corresponding to
large wavelengths $2\pi/q$).  Already in 1931, Onsager explained how
the microscopic equations in the limit of $q\to0$ and small frequency
lead to the hydrodynamic equations \cite{Onsager31}. He suggested that
in an infinite system a perturbation described by macroscopic
hydrodynamic equations decays from its inital value according to the
identical dynamical equations as a long--wavelength and small
frequency fluctuation around local thermodynamic equilibrium
\cite{Kadanoff63}.  As an important side-product, this correlation
functions approach has led to general and exact microscopic
expressions for the phenomenological transport coefficients of the
hydrodynamic equations (the Green--Kubo relations).

Hydrodynamic equations, which are partial differential equations,
require temporal and spatial boundary conditions to give unique
solutions; see e.g.\ the examples in \cite{Landau59}.  Following
Onsager, only the former are understood from microscopic many--body
approaches, while, by studying infinite systems, the latter have been
neglected. Within the phenomenological approach, simple continuity
considerations lead to the required conditions on surfaces, yet their
derivation from information about microscopic interactions and
molecular parameters still appears desirable. First, this would
provide rigorous statistical mechanics definitions of the parameters
characterising the boundary condition; second, different conditions
(like stick or slip for fluid flow, or flux versus no--flux with or
without adsorption of particles at a surface) could be predicted from
molecular interactions; and third, generalizations beyond the true
hydrodynamic limit (e.g.\ for finite geometries) would become
possible. Only the question about the tangential velocity of a flow
along a solid boundary has a long history, which goes back to Maxwell,
and, for rarefied gases, is quite well answered in the framework of
the Boltzmann equation (Knudsen-layer problem \cite{Cercignani}).
Yet, beyond the dilute limit a fluctuating hydrodynamics calculation
by Wolynes \cite{Wolynes76} has uncovered the subleties arising from
back flow patterns (coupling of hydrodynamic modes), and only rather
recently there has been the first study of this problem in the
microscopic correlation functions approach by Bocquet and Barrat
\cite{Bocquet94}. In our study of the simpler system of a single
conserved variable which macroscopically obeys a diffusion equation,
we follow the approach of Bocquet and Barrat and connect a microscopic
linear response calculation to the macroscopic hydrodynamic
description via a generalization of Onsager's regression hypothesis.

On the macroscopic level, the number density $n(\vec r,t)$ of
interacting Brownian particles at the space point $\vec r$ and at time
$t$ obeys a diffusion equation
\begin{equation}\label{e1}
 \partial_t\, n(\vec r,t) = D \; \nabla^2\; n(\vec r,t) \; ,
\end{equation}
where $D$ is the (gradient) diffusion coefficient which enters \gl{e1}
as a phenomenological transport coefficient;
$\partial_t=\partial/\partial t$ denotes a partial time derivative and
$\nabla=\partial/\partial\vec r$ a spatial gradient. If the diffusing
particles border a solid surface which moves with velocity $\vec v(t)$
then the number of particles (per unit area and time) displaced by the
surface, $n \vec v$, needs to be balanced by a particle flux $\vec j$
away from the boundary. As the latter obeys $\vec j = - D \nabla n$,
the no--influx boundary condition on the solid surface becomes
\cite{Landau59}:
\begin{equation}\label{e2}
\uvec e_n \left[ n(\vec r,t) \; \vec v(t) + D\, \nabla n(\vec 
r,t) \right]_{\rm bd}  = 0 \; ,
\end{equation}
here $\uvec e_n$ is a unit vector normal to the surface, whose
position is abbreviated as ``${\rm bd}$'' for boundary.

In sections 2. and 3.~of this manuscript, \gl{e2} will be derived up
to linear order in $\vec v$ by coarse graining the many--body
statistical mechanics description of interacting Brownian
particles. At first the appropriate microscopic kernel is found
(sect.~2.), and then its small $q$ and $\omega$ behavior discussed
(sect.~3.). The calculation entails the (standard) derivation of
\gl{e1} including the Green--Kubo type calculation of $D$. The solution
of \gls{e1}{e2} around a spherical object to linear order in the
perturbing velocity $v$ is summarized in Appendix A for comparison
reasons, while Appendix B contains technical material. Section
4.~describes an application of our results.  The power--law decay of
the force experienced by a large sphere moving among the Brownian
particles is deduced along with its mean-squared displacement, which
exhibits a long time tail.

\section{Microscopic approach}

\subsection{Smoluchowski equation, and notation}

The statistical mechanics basis for interacting Brownian particles is
given by the Smoluchowski equation, which is a generalized diffusion
equation in high-dimensional phase space \cite{dhont,Pusey85}. It
describes the temporal evolution of the many--body probability
distribution $\Psi(\{\vec r_i\},t)$, which depends on the positions
$\vec r_i$ of all particles, $i=1,\ldots,N+1$, where we consider $N$
identical bath particles with Brownian diffusion coefficient $D_i=D_0$
and one additional tracer with index $s=N+1$ and diffusion coefficient
$D_s$:
\begin{equation}\label{mi1}
\partial_t \Psi =  \sum_i D_i\; \partial_i
\cdot \left(\partial_i-\vec F_i\right)\; \Psi \; .
\end{equation}
Here $\partial_i=\partial/\partial\vec r_i$, and energies are measured
in units of the thermal energy.  The particle interactions enter
\gl{mi1} via the potential forces $\vec F_j\equiv-\partial_j V(\{\vec
r_i\})$ resulting from interactions between the bath particles ($V^p$)
and between tracer and bath particles ($V^s$), $V=V^p+V^s$. Dynamic
effects due to the background medium (hydrodynamic interactions) are
neglected at the present stage. To reach the hydrodynamic limit we
will take the size of the tracer to become much larger than the bath
particle size. For the presentation in the main text, the tracer is
assumed to be immobile from the outset, $D_s=0$, and Appendix B
verifies that the limit $D_s\to0$ can be taken after the formal
manipulations. Summation and indices are from now on always understood
to run from 1 to $N$, i.e., to exclude the index $s=N+1$ for the
tracer. To simplify the presentation, we also introduce the radii
$a_s$ and $a$ of the tracer and the bath particles, respectively. It
is however important to realize that we consider arbitrary isotropic
short ranged particle interactions, where $a$ and $a_s$ may be
effective state dependent sizes as e.g.\ in the case of soft
repulsions of the form $V^p(r=|\vec r_i-\vec r_j|)\propto r^{-12}$.

For the following, we introduce some further notational
conventions. It is convenient to work with the backward or adjoint
Smoluchowski operator
\begin{equation}\label{mi2}
\Omega\equiv D_0 \; \sum_{i}\, \left(\partial_i+\vec
F_i\right)\cdot\partial_i\; ,
\end{equation}
which gives the time evolution of variables $A(\{\vec r_i\})$ on phase
space: $\partial_t A = \Omega A$. It also determines the time
evolution of correlation (fluctuation) functions $\Phi_{AB}(t)=
\langle A^* \exp{\{\Omega t\}} B \rangle/ \langle A^*B\rangle$, which
we normalize by their equal time values calculated by canonical
averaging with the Gibbs--Boltzmann weight $\langle \ldots\rangle
\propto \int \prod_{i=1}^{N+1} d\vec r_i\; \ldots e^{-V}$. Note that here the
tracer-particle interactions enter in full non--linear detail, and that
the equilibrium weight is a stationary solution of \gl{mi1}.

The fluctuating microscopic bath particle density at position $\vec r$
is given by $\rho(\vec r) = \sum_j\, \delta(\vec r - \vec r_j)$,
with spatial Fourier transform, $\varrho_{\vec q} = \sum_j\,
\exp{\{i\vec q \vec r_j\}}$, where the $q=0$ contribution from the
constant bulk density $n$ will be neglected. The corresponding tracer
density fluctuation is given by $\rho^s_{\vec q} = \exp{\{i\vec q \vec
r_s\}}$.  Temporal Fourier decomposition shall be denoted by $A_\omega
= \int_{-\infty}^\infty dt\; e^{i\omega t} A(t)$, while the Laplace
transformation is used with the convention: $A(\omega)=\int_0^\infty dt\;
e^{i\omega t} A(t)$.

\subsection{Generalized Onsager regression hypothesis}

The connection between the statistical mechanics description on the
Smoluchowski level and the macroscopic hydrodynamic picture shall be
made by comparing the density fluctuations predicted from both
descriptions for the identical given boundary
problem in a simple geometry.

In order to use the familiar Smoluchowski operator of \gl{mi1}, we
consider the motion of particles around a spherical object, the
tracer.  Bath particle $j$ experiences the short-ranged force
$F^s_j=-\partial_j V^s$ close to it. Moving the tracer, by unspecified
external means, with velocity $\vec v(t)$ induces a particle flux at
its surface, which in linear order in v equals $\vec j^{\rm bd}(t) = n
\vec v(t)$ on the macroscopic level. Deviations of the coarse grained
particle density around the tracer $n(\vec r,t)$ from the bulk value
$n$ would enter in higher order in v only.  The disturbance on the
microscopic level, required to induce this applied particle flux thus
can be obtained from requiring the non--equilibrium average of the
tracer velocity to agree with the macroscopic value up to non--linear
corrections:
\begin{equation}\label{mi3}
\langle \partial_t\,
\vec r_s \rangle^{\rm (ne)} \; = \; \vec v(t) + {\cal
O}({\rm v}^2)\; .
\end{equation}

Adiabatically turning on the applied velocity in the infinite past
eliminates initial value contributions in the deviatoric density,
$\delta n(\vec r,t)=n(\vec r,t)-n$, and allows us to use a Fourier
decomposition, $\vec v(t) = \int (d\omega/2\pi)\; e^{-i\omega t} \vec
v_\omega$. Linear response theory then connects the density deviation
to the given disturbance via a (vector) susceptibility $\vec \chi(\vec
r,t)$. It vanishes for $t<0$ because of causality, and its spatial
argument $\vec r$ is measured from the tracer sphere center. After
Fourier transformation, both the macroscopic hydrodynamic result
(cf.\ Appendix A) and the microscopic result (cf.\ sections 2.3 and 2.4)
can be written as
\begin{equation}\label{mi4}
\delta n_{\vec q,\omega} \; =\;\; n\; \vec v_\omega \cdot \vec
\chi_{\vec q}(\omega)+  {\cal
O}({\rm v}_\omega^2)\; .
\end{equation}

Now, in the spirit of Onsager's hypothesis we assume that the
microscopic calculation of \gl{mi4} reduces to the macroscopic
solution for smooth and slow fluctuations, viz. in the limit of small
frequencies and wavevectors. Yet, in order to derive hydrodynamic
boundary conditions, the coarse graining must be taken with respect to
the bath particle size $a$ only, while the tracer size is required to
satisfy $a_s\gg a$.  Thus we keep $a_s$ fixed so that the macroscopic
diffusion equation description, while it applies for distances $r \gg
a$ only, nevertheless includes both far field ($r \gg a_s$) and near
field ( $r \ll a_s$). The latter case is equivalent to considering the
density profile $\delta n(z,t)$ at a distance $z$ close to a planar
wall obtained formally when taking $a_s\to\infty$. Although this
limit does not provide a faithful representation of a macroscopic
boundary as an assembly of atoms, it has the virtue of being the
conceptually simplest realization of a hydrodynamic boundary problem.
 
Two aspects of the described approach are worth mentioning: First,
while \gls{mi3}{mi4} are linear in the applied boundary flux, the
particle--wall (tracer) interactions are included exactly. Thus on a
local length scale the unperturbed equilibrium density variations
arise, which somewhat differs from the approach to shear flow past a
surface in Ref. \cite{Bocquet94}.  Second, as discussed e.g.\ by
Kadanoff and Martin for the initial value problem \cite{Kadanoff63}, a
general perturbation to the fluid induces fluctuations in the
non--conserved variables, which have to die out before the hydrodynamic
description applies. For the present boundary perturbation the same
reasoning applies, and thus the hydrodynamic description only holds
for large distances, while locally deviations from the hydrodynamic
solution need to appear; for rarefied gases these Knudsen-layers
effects are familiar \cite{Cercignani}. In the present many--body
linear response calculation the technical difficulty is connected to
coarse graining across the equivalent layer, which has a width
connected to the particle size $a$.

\subsection{Linear response calculation}\label{sec:lr}

In order to proceed, the perturbation to the Smoluchowski operator
$\Omega$ needs to be found which gives the required velocity of the
tracer in \gl{mi3}. Without hydrodynamic interactions, the
perturbation equivalent to a constant solvent velocity, which is felt
solely by the tracer, is by inspection
\begin{equation}\label{mi5}
\Delta \Omega = \vec v(t) \cdot \partial_s \; .
\end{equation}
A standard linear response calculation using its adjoint $\Delta
\Omega^\dag = - \vec v(t) \cdot \partial_s$, which acts on the
probability density  in  \gl{mi1} \cite{dhont}, gives the
resulting deviation in an arbitrary variable $A$
\begin{equation}\label{mi6}
\langle \delta A(t) \rangle^{\rm (ne)} \equiv
\langle A(t) \rangle^{\rm (ne)} -
\langle A(t) \rangle 
\;  =\; -
\int_{-\infty}^t\!\!\! d\tau\; \vec v(\tau) \cdot \langle \vec F_s\;
e^{\Omega (t-\tau)}  \; A \rangle \; + {\cal O}({\rm v}^2)\; .
\end{equation}
Thus, \gl{mi3} becomes
\begin{equation}\label{mi3_2}
\langle \partial_t\, \vec r_s \rangle^{\rm (ne)} = 
\langle (\Omega + \Delta \Omega)\;  \vec r_s \rangle^{\rm (ne)} \;  = \;
\langle  \vec v(t) \cdot \partial_s
\vec r_s \rangle + {\cal O}(D_s/D_0,{\rm v}^2)\; .
\end{equation}
See Appendix B, for a more careful discussion for finite tracer
diffusivities $D_s>0$.  As required, the perturbation \gl{mi5} gives
the average velocity of the tracer, which enters the macroscopic
boundary condition \gl{e2}.

The linear response formula can also be applied to the microscopic
density field $\rho(\vec r')$ at a (vector) distance $\vec r$ from the
tracer center: $\vec r'=\vec r_s + \vec r$. Its unperturbed
equilibrium value is proportional to the familiar tracer-particle pair
correlation function \cite{hansen}: $g^s(\vec r) = (1/n) \sum_{i}
\langle \delta\left[\vec r - (\vec r_i - \vec r_s) \right] \rangle$,
which gives the probability of finding bath particles at a distance
$r$ from the center of the tracer.    The linear deviation
in the density around the tracer induced by the perturbation, \gl{mi5},
follows immediately from \gl{mi6}, and by comparison with \gl{mi4},
the required linear response susceptibility is found:
\begin{eqnarray}\label{mi8}
\vec \chi_{\vec q}(\omega) &=&  -\frac{1}{n} \;
\langle \vec F_s \; \frac{-1}{\Omega + i \omega} \; \rho^{s*}_{\vec q} \,
\rho_{\vec q} \rangle \; ,
\quad \mbox{ or}\nonumber\\ 
\vec \chi(\vec r,t) &=&  -\frac{1}{n} \;
\langle \vec F_s \; e^{\Omega t} \; \rho(\vec r +
\vec r_s) \rangle \; \theta(t) \;,
\end{eqnarray}
where the step-function $\theta(t)$ expresses causality.

\subsection{Time scale separation}\label{sec:tss}

The exact linear response susceptibility varies on rapid microscopic
time and length scales but also on smooth and slow ones, which are
amenable to a hydrodynamic description.  The Zwanzig-Mori projection
operator formalism enables one to disentangle these contributions by
splitting the resolvent into fast and slow subspaces \cite{hansen}.
The resolvent $R(\omega) = \frac{-1}{\Omega+i\omega}$ arises in the
Laplace transform of a general correlation function,
$\Phi_{AB}(\omega) = \langle A^* R(\omega) B \rangle/\langle A^*
B\rangle$, and contains poles which shift to vanishing frequency for
smooth fluctuations ($q\to0$). 
These so-called hydrodynamic poles are connected
with the exact conservation laws of the system, and the Zwanzig-Mori
formalism isolates them.  In the present situation, where \gl{mi1}
holds, there are only poles connected with particle number
conservation: each particle, including the tracer, is conserved as is
the total density, $\partial_t \rho_{\vec q} \propto q$ for $q\to0$.
While the more careful calculation in Appendix B takes into account
the tracer, here we chose for the slow subspace the one spanned by the
total density only. That is, we use the projector $P = \rho_{\vec q}
\rangle (N S_q )^{-1} \langle \rho^*_{\vec q}$, which is normalized by
the equilibrium Brownian particle structure factor $S_q = \langle
\rho_{\vec q}^* \rho_{\vec q}\rangle / N$ \cite{hansen}.  The
justification for this simplification is provided by the thermodynamic
limit, in which only a non--extensive number of particles actually
interacts with the tracer; see below and Appendix B.
 
The exact identity obtained in the Zwanzig--Mori projection operator
formalism \cite{Goetze89b} gives for a general
fluctuation function:
\begin{equation}\label{mi9}
\avg{A^* R(\omega) B}=\avg{A^* R'(\omega) B}+\avg{A^* (1+R'(\omega)
\Omega) P R(\omega) P (1+ \Omega R'(\omega) ) B}\; ,
\end{equation}
where the reduced resolvent describes the fast dynamics decoupled
from the slow fluctuations of the conserved density:
\begin{equation}\label{mi10}
R'(\omega) = Q \frac{-1}{Q \Omega Q + i \omega } Q\quad\mbox{with }
Q=1-P\; .
\end{equation}
Thus, the coupling of the arbitrary variables $A$, $B$ to the slow
conserved density is found; explicitly it is obtained when writing out
$P R(\omega) P$ in \gl{mi9}, and the slow variable couples in with
static (i.e.\ $\langle A^* \rho_{\vec q}\rangle$) and frequency
dependent (i.e.\ $\langle A^* R'(\omega) \Omega \rho_{\vec q}\rangle$)
overlaps.

In order to apply \gl{mi9} to the correlation function in \gl{mi8},
the (expected) problem arises that it is formed with variables that
are not defined in a translationally invariant manner. Translational
symmetry is broken by the boundary (i.e. measuring distances from the
tracer). On the macroscopic level this could be handled by introducing
the appropriate eigenfunctions that satisfy the boundary conditions
for the prescribed geometry. Yet, on the microscopic level this would
require determining the many--body eigenfunctions of the Smoluchowski
operator \gl{mi2} for a given force field arising from $V^s$. Within
the framework of fluctuating hydrodynamics Wolynes achieved a related
task in a scattering--formalism calculation for the flow of a Newtonian
fluid past a wall \cite{Wolynes76}. His calculation focused on the
non--linear coupling of the hydrodynamic modes and thus could
circumvent the study of the local variables close to the
boundary. Consequently, he did not determine the boundary position
microscopically and instead introduced a short distance cut--off (in
his case irrelevant). Because we aim for an exact determination of the
boundary condition, we chose the plane--wave decomposition of the
density fluctuations which enables one to use \gl{mi9}, and apply it
to the resolvent in a shifted coordinate system
\begin{equation}\label{mi11}
R_{\vec q}(t) = \rho^s_{\vec q}\; R(t) \; \rho^{s*}_{\vec q} \;
 = \; R(t) \; ( 1 + {\cal O}(D_s/D_0) ) \;.
\end{equation}
It agrees with the original resolvent only if thermal tracer
fluctuations are neglected (cf.\ Appendix B). In this limit, the
Fourier transformed susceptibility becomes
\begin{equation}\label{mi12}
- n\; \vec \chi_{\vec q}(\omega) = 
\langle \vec F^{s*}_{\vec q}\;  R(\omega) \;
 \rho_{\vec q} \rangle =
\langle \vec F^{s*}_{\vec q} \, \left[ 1 + R'(\omega)\, \Omega
\right]\, \rho_{\vec q} \rangle \; \Phi_{\vec q}(\omega)\; ,
\end{equation}
where we have introduced the tagged force density fluctuation $\vec
F^s_{\vec q}=\vec F_s\rho^{s}_{\vec q}$ and the (normalized) density
correlator $\Phi_{\vec q}(\omega)= \langle \rho_{\vec q}^* R(\omega)
\rho_{\vec q}\rangle/(NS_q)$. Application of \gl{mi9} to the latter
gives the familiar expression
\begin{equation}\label{mi13}
\Phi_{\vec q}(\omega) = \left[ - i \omega - \frac{\langle \rho_{\vec
q}^*\Omega \rho_{\vec q}\rangle}{NS_q} + \frac{\langle \rho_{\vec
q}^*\Omega R'(\omega) \Omega \rho_{\vec q}\rangle}{NS_q}
\right]^{-1}
\to  \left[ - i \omega + q^2 \frac{D_0}{S_0} \right]^{-1} \; .
\end{equation}
The second expression in \gl{mi13} is taken in the hydrodynamic limit
of small frequencies and wavevectors, where it gives the (transformed)
fundamental solution of the diffusion equation \gl{e1}. This leads to
the known microscopic definition of the gradient diffusion
coefficient, $D=D_0/S_0$.  Here, $S_0$ is a normalized
compressibility. The result for $D$ may be called of Green--Kubo type
because its apparent static nature originates in an instantaneously
decaying associated current.

The frequency independent (or instantaneous) overlap in \gl{mi12} can
be expressed in terms of the Fourier transform of the non--trivial
part, $h^s(\vec r)=g^s(\vec r)-1$, of the tracer-particle pair
correlation function introduced above, $\langle \vec F^{s*}_{\vec q}
\rho_{\vec q} \rangle = i \vec q\, n h^s_{\vec q}$, but little further
simplification is possible in the retarded second term. Upon
introducing the total force density fluctuation $\vec F_{\vec q} =
\sum_j \vec F_j \, e^{i \vec q \vec r_j}$, $Q \Omega \rho_{\vec q} =
i Q \sum_j \vec q \cdot \vec F_{\vec q}$, and the final (still exact)
result for $\vec \chi$ becomes (for $t>0$)
\begin{equation}\label{mi14}
\vec \chi_{\vec q}(t) = - i \vec q\, h^s_{\vec q} \; \Phi_q(t) - i
\frac{D_0}{n} \int_0^t\!\!\!d\tau\; \langle \vec F^{s*}_{\vec q}\;
R'(t-\tau)\; \vec q \cdot \vec F_{\vec q} \rangle \; \Phi_q(\tau) \; .
\end{equation}
It is written as function of time to clearly present the instantaneous
(first term) and retarded coupling of the density fluctuations to the
susceptibility. Because of Newton's third law, the potential force
$\vec F_s$ felt by the tracer can be reexpressed as the negative of
the total force exerted by the tracer on all particles; $\vec F_s = -
\partial_s V^s = \sum_i \partial_i V^s\equiv-\vec F_0$.  For the same
reason it is opposite equal to the integrated total force; $\vec F_s
=\vec F^s_{\vec q=0} = -\vec F_{\vec q=0}=\sum_i\partial_iV_i=-\vec F_0$.  A
noteworthy aspect of the (straightforward) calculation in this section
concerns the thermodynamic limit which is required in order for the
obtained bulk quantities to take their standard values for an
unbounded system.  For example, the tracer bath particle interactions
enter the expression for $S_q$ (and consequently for $D$) via the
equilibrium distribution function. Nevertheless, in the thermodynamic
limit this correction vanishes because the assumed short-ranged
interaction of the bath particles with the tracer decays beyond the
distance of a few $a$, and the bulk of the particles is not affected.

\section{Coarse graining and discussion}

The exact correlation function, \gl{mi14}, describes the response of
the system to an injected boundary flux of particles on all length
scales. In order to derive the hydrodynamic boundary condition, coarse
graining is required as discussed in section 2.2. Appendix A
collects the results from the macroscopic approach in order to compare
them with the small wavevector and frequency limits of the microscopic
susceptibility.

\subsection{Instantaneous response}

In order to familiarize oneself with \gl{mi14}, it is useful to
consider a rapid velocity pulse on the tracer at time $t_0$, $\vec
v(t) = \vec V \delta(t-t_0)$, and to concentrate on the instantaneous
response of the density:
\[
\delta n(\vec r,t=t_0) = n \, \vec V \cdot\vec \chi(\vec r,t=0)
=  n \, \vec V \cdot \nabla h^s(\vec r)\; .
\]
This arises from the first term in \gl{mi14}, which simplifies because
of $\Phi_{\vec q}(t=0)=1$, and is determined by the equilibrium
density profile of bath particles around the tracer.  The inserted
flux, $n \vec V$, is packed close to the boundary according to the
equilibrium fluid stucture $h^s$ of the bath particles.  The (Ursell)
function $h^s$ varies between the universal limits, $h^s=-1$ for
short distances where the hard core volumes of the particles are
excluded by the tracer, and $h^s=0$ far away from the tracer.
In between, it shows layering over a distance of the order of a few
$a$.

In the hydrodynamic limit, which corresponds to $a_s \gg a$ here, the
present work provides the connection of the position of the boundary
to the molecular interaction potential $V^s$. At a radial distance
$\sigma$ from the tracer center, $h^s$ varies rapidly \cite{hansen}
and in the limit $a\to0$ with fixed $a_s$ and $r$, may loosely be
taken as a step function, $h^s(\vec r)=-\theta(\sigma-r)$.  The
macroscopic sphere asymptotically becomes impenetrable for the bath
particles irrespective of the exact interaction potential. The latter
however determines the exact boundary position $\sigma$, and its
definition becomes:
\begin{equation}\label{mi15}
h^s_{\vec q} \to - 2 \pi \sigma^3 \; f(q\sigma) +
{\cal O}(\sigma^2) \for a_s\gg a \;\mbox{
and } q a \ll 1 \;,
\end{equation}
where $f(x)=(\sin x-x\cos x)/x^3$. Whenever \gl{mi15} does not hold,
possibly for long ranged forces or wetting situations, we expect
\gl{e2} to be violated. Such situations are excluded in the following.
A finite (positive or negative) surface excess density enters in the
corrections of order $\sigma^2$. In the following sections we show
that exactly the same structure also appears in the retarded
contributions to \gl{mi14}, and that the boundary position $\sigma$
thus is a static equilibrium concept (see however
Ref. \cite{Bocquet94} for Newtonian fluid flow).

In the limit of $a_s\gg a$, the rapid variation of $h^s$ can be used
to define a one-dimensional cut through the density profile, which in
the limit $a_s\to\infty$ (and consequently $\sigma\to\infty$) would
correspond to the situation at a wall \cite{hansen}.  With the wall at
$x=\sigma$, and its normal vector pointing along $\uvec x$, the wall
profile $h^{s W}$ as function of $\bar{\vec r}=\vec r-\sigma \uvec x$,
$\bar x= \bar{\vec r}\cdot\uvec x$ follows
\begin{equation}\label{mi16}
h^s(\vec r) \to h^{s W}(\bar x) + {\cal O}(\bar r/\sigma) \for
\sigma\to\infty \; .
\end{equation}
It obeys, $h^{s W}(\bar x\to-\infty)\to - 1$ and $h^{s W}(\bar x\to
\infty)\to 0$, with rapid variations on a length scale of order $a$
around $\bar x\approx 0$.  Its one-dimensional Fourier transform is
given by
\begin{equation}\label{mi17}
h^{s W}_{q_x} = \int_{-\infty}^\infty\!\! d\bar x\; e^{iq_x \bar x}\;
h^{s W}(\bar x) = \frac{{\cal H}_{q_x}}{-iq_x} \; =\; \frac{1}{-iq_x}
+ {\cal H}' ( 1 + {\cal O}(q_x a) )\; ,
\end{equation}
where the constant ${\cal H}'$ is the surface density excess divided
by $n$ and is of order $a$ itself. By shifting the origin to $\bar
x=0$ ($x=\sigma$), we eliminated (for simplicity) a phase factor
$e^{iq_x \sigma}$ in $h^{s W}_{q_x}$, which would prove convenient
when keeping track of the wall position.

\subsection{Near--field solution}\label{sec:nfs}

Generically, boundary conditions are formulated when considering the
motion in a half space bounded by a planar surface (wall).  As
discussed above for the instantaneous response, this situation can be
realized in \gl{mi14} by taking the limit $a_s\to\infty$ and
calculating $\vec \chi(\vec r=\sigma \uvec x+\bar{\vec
r},t)=\chi^{W}(\bar x,t) \uvec x +{\cal O}(\bar r/\sigma)$ to
non--vanishing order. The result $\chi^{W}(\bar x,t)$ describes the
motion close to an infinite plane wall or, in general, the near--field
solution for non--planar solid surfaces. Only its small wavevector
limit is required in the following, and this simplifies because the
force exerted on the diffusing particles by the wall (for
$a_s\to\infty$) is perpendicular to it:
\begin{equation}\label{mi18}
\chi^{W}_{q_x}(\omega) \sim  \left[  1 
+ i q_x\; \frac{D_0}{n}  \; \langle \uvec x\cdot\vec F_{0}
\; R(\omega)\;  \uvec x\cdot\vec F_{0} \rangle \; \right]
\Phi_{q_x}(\omega) \; .
\end{equation}
Here we have used that for vanishing wavevector the reduced resolvent
in the relaxation kernel again agrees with the full dynamics \cite{hansen}.
The retardation kernel in \gl{mi18} therefore has the familiar
Green--Kubo form. If it could be replaced by a constant rate for small
frequency, $\langle \uvec x\cdot\vec F_{0}
\; R(\omega\to0)\;  \uvec x\cdot\vec F_{0} \rangle \to
\Gamma$, then for consistency the square bracket would become $\left[
\ldots \right]\to 1 +{\cal O}(q)$. Fortunately, in an exact
calculation for vanishing concentration of hard Brownian spheres,
$n\to0$, Dieterich and Peschel have shown that \cite{Dieterich79}
\begin{equation}\label{mi19} 
i \frac{D_0}{n}\; \langle \uvec x\cdot\vec F_{0}
\; R(\omega)\;  \uvec x\cdot\vec F_{0} \rangle 
=  \sqrt{\frac{-iD_0}{\omega}} \;\;
\left( 1 + {\cal O}(n a^3) \right)\; .
\end{equation}
In this limit $D=D_0$, and the result of
Eqs. (\ref{mi4},\ref{mi13},\ref{mi18}) agrees with the solution of the
hydrodynamic equations, \gls{e1}{e2}, in the considered geometry; see
\gl{a4}. This proves the boundary condition \gl{e2} in the dilute
limit of hard spheres \cite{Shapiro90}.

For the general situation of interacting Brownian particles at finite
concentrations, no exact calculations of the relaxation kernel in
\gl{mi18} are known.  We proceed by performing a mode coupling
approximation \cite{Pusey85,Hess83,Goetze91b}, where the starting
point is the more general expression of \gl{mi14} as it captures
near-- and far--field terms. The conserved density fields are the slow
variables and in the lowest pair--fluctuation approximation the
overlap of the fluctuating forces with $\rho_{\vec k}\rho^s_{\vec k'}$
needs to be considered.\footnote[5]{It can be expected  that this
approximation does not give numerically exact hydrodynamic results
\cite{keyes-masters85}, but Schofield and Oppenheim \cite{Schofield92}
have argued that this problem can be overcome by systematically taking
the overlap with higher order density products into account.} In the
small wavevector limit of interest, the memory function becomes
identical to the well studied tracer force autocorrelation kernel, and
its mode coupling result can be taken from the literature
\cite{Goetze91b}.
\begin{equation}\label{mi20}
\frac{D_0}{n}\; \langle \vec F_{0}
\; R'(t)\;  \vec q\cdot\vec F_{0} \rangle 
\approx \; D_0\; \vec q \cdot \lim_{q\to0}\;
\int\!\!\frac{d^3k}{(2\pi)^3}\;  (\vec k\vec k)\; \frac{h^s_{\vec
k}h^s_{\vec q-\vec k}}{S_k}\; \Phi_{\vec k}(t) \; ,  
\end{equation}
where the tracer density fluctuation function does not appear
explicitly (it equals 1 because of $D_s=0$).

The mode coupling approximation of the relaxation kernel can be
applied to the wall or near--field problem upon the realization that
the forces arise from density fluctuations whose probability depends
on the wall distance according to $h^{s W}(\bar x)$ and is independent
of the parallel coordinates, $\bar y$ and $\bar z$. For the relevant
wavevector region\footnote[6]{First, Eq. (\protect\ref{mi20}) is
transformed to $r$-space and then $h^s(\vec r)$ is analyzed according
to Eq. (\protect\ref{mi16}). The procedure is checked a posteriori
from the convergence of the integral.}, the $h^s_{\vec k}$ in
\gl{mi20} thus corresponds to
\begin{equation}\label{mi21}
h^s_{\vec q} \approx h^{sW}_{\vec q} = (2\pi)^2 \; \delta_\|(\vec q)\;
\frac{{\cal H}^s_{q_x}}{-i q_x}\; ,
\end{equation}
where $\delta_\|(\vec q)$ restricts the parallel wavevector to vanish,
$q_y=q_z=0$, and the wall profile function ${\cal H}^s$ defined in
\gl{mi17} enters.  We find the approximation:
\begin{equation}\label{mi22}
\frac{D_0}{n}\; \langle \uvec x\cdot\vec F_{0}
\; R(\omega)\;  \uvec x\cdot\vec F_{0} \rangle 
\approx   (2\pi)^2 \; \delta_\|(\vec q) \; \int_{-\infty}^\infty\!\!
\frac{dk_x}{2\pi}\; \frac{|{\cal H}^s_{k_x}|^2\; D_0}{S_{k_x}} \;
\Phi_{k_x}(\omega)\; ,
\end{equation}
where in the integrand $k_y=k_z=0$.  Importantly, in the hydrodynamic
limit $\omega\to0$, the integral converges already for such small
$k_x\ll 1/a$ that the structure functions can be replaced by their
homogeneous zero wavevector limits, ${\cal H}^s_{k_x}\to1$ and
$S_{k_x}\to S_0$. The latter is the bulk compressibility required to
turn the single particle $D_0$ into the gradient diffusion constant
$D$.  It is thus interesting that the exact one-dimensional result of
\gl{mi19} in the dilute limit only applies to the present case because
the (isolated) particles experience no interactions and thus $D=D_0$
in \gl{mi19}. Because the integration in \gl{mi22} is dominated by
$k\ll a$, the density correlator can be replaced by its universal
hydrodynamic limit from \gl{mi13}.

Collecting all terms together and performing the integrations
gives the mode coupling approximation
\begin{equation}\label{mi23}
\vec \chi_{\vec q}(\omega) \approx \uvec x  \;
(2\pi)^2 \; \delta_\|(\vec q)\; \left\{  1 
+ \sqrt{\frac{-iD q_x^2}{\omega}} \left[ \frac 12  + \ldots \right]
\right\}
\; \frac{1}{D\, q_x^2 - i \omega } \; ,
\end{equation}
where the term $\frac 12$ in the square bracket arises from the
considered pair density projections and $\dots$ indicate higher order
density projections which should be taken into account following an
expansion procedure developed by Schofield and Oppenheim
\cite{Schofield92}.  Assuming the series of density projections in the
square bracket in \gl{mi23} to sum up to one, the comparison with
\gl{a4} proves the correctness of the boundary condition \gl{e2}, now
at finite concentrations.

\subsection{Far--field solution}

While the verification of the boundary condition \gl{e2} is achieved
by the  calculation of the near--field, it is
instructive to also consider the density fluctuations very far away
from the spherical boundary. In this limit, both particle sizes $a$
and $a_s$ are small compared to the wavelength and the susceptibility in
\gl{mi14} simplifies to 
\begin{equation}\label{mi24}
\vec \chi_{\vec q}(t) \sim -i\vec q\; \alpha \; \Phi_q(t)\with
\alpha = h^s_0 - \frac{D_0}{3n}\; \int_0^\infty\!\!  dt\; \langle \vec
F_{0}(t) \cdot \vec F_{0} \rangle \; ,
\end{equation}
for $qa_s\ll1$; where the density correlator takes its hydrodynamic form,
$\Phi_q(t)=\exp{(-q^2Dt)}$ from \gl{mi13}. The density profile around
a moving tracer rearranges by particle diffusion and thus requires
more and more time the larger the involved distances. In the steady
case, $\omega=0$, a power law density profile develops
\begin{equation}\label{mi25}
\delta n_{\omega=0}(\vec r) \sim -
\frac{\alpha n}{4\pi D} \; \frac{\vec v_{\omega=0} \cdot \uvec r}{r^2} \; ,
\end{equation}
as follows from Eqs. (\ref{mi4},\ref{mi13},\ref{mi24}) and
transformation to $\vec r$-space. It is of similar nature to the Oseen
velocity profile around a colloidal particle in a Newtonian solvent
\cite{Landau59}, as such long--ranged patterns generally arise in
hydrodynamic steady states \cite{Machta82}. In the Brownian particle
context, it is well known from calculations in the dilute limit,
$n\to0$ \cite{dhont,Ackerson82,Cichocki92}, and the amplitude factor
$\alpha$ extends those calculations to finite densities.

Interestingly, the expression for $\alpha$ in \gl{mi24} holds for
arbitrary size ratios $a_s/a$, even beyond the macroscopic
hydrodynamic limit, which is obtained for $a_s\gg a$. For dilute hard
spheres, the known result $\alpha(n\to0,a/a_s\to0)=-2\pi\sigma^3$
(where $\sigma=a_s$) \cite{Ackerson82,dhont} agrees with the
expectation from the macroscopic calculation, \gl{a3}. For finite
densities the mode coupling approximation \gl{mi20} can be used
\begin{equation}\label{mi26}
\alpha\approx h^s_0 - \lim_{\omega\to0} \; D_0 \;
\int_0^\infty\!\!\frac{dk}{6\pi^2}\; \frac{(k^2 h^s_k)^2}{S_k} \;
\Phi_k(\omega)\; .
\end{equation}
In the limit of a macroscopic tracer, which becomes impenetrable
to the Brownian particles so that $h^s(r)$ approaches a step function
as argued in \gl{mi15}, the integration in \gl{mi26} already
converges for $ka\ll1$. The density correlator is then given by the
hydrodynamic limit and the structure factor equals the
compressibility, $S_k=S_0$, so that
\begin{equation}\label{mi27}
\alpha \approx - 2 \pi \sigma^3 \left\{ \frac 23 + \left[ \frac 29 +
\ldots \right] \right\} \; .
\end{equation}
The leading contribution $2/3$ arises from the static term, while we
may again expect \cite{Schofield92} the Green--Kubo expression to sum
up to the missing $1/3$ if, extending our pair--density factorization,
higher order density fluctuations are included.

\section{Application to diffusive long time tail}

An immediate consequence of the long--ranged structure built by
particle diffusion around a macroscopic tracer, are slow time
dependent fluctuations in the force the tracer feels.  An interesting
aspect of these so-called long time tails is that hydrodynamic
calculations provide insights into them \cite{Landau59,Zwanzig70},
even in the presence of boundaries \cite{Hagen97}. As an application
of the above discussion of boundary conditions for the diffusion
equation, we study the long time tail in the force autocorrelation
function of a tracer diffusing among Brownian particles.  This extends
the knowledge available at infinite dilution
\cite{dhont,Ackerson82,Cichocki92}.

As a first step, the constitutive equation connecting the force a
particle feels to the fluctuations 
of the conserved variable, the bath density, is required. It follows
from the Zwanzig--Mori 
decomposition  as \cite{Goetze89b} 
\begin{equation}\label{ltt1}
\langle \delta \vec F^{s*}_{\vec q}(\omega) \rangle = \frac{\langle
\delta \rho^*_{\vec q}(\omega) \rangle }{NS_q} \left( \langle \vec
F^{s*}_{\vec q} \rho_{\vec q} \rangle + \langle \vec F^{s*}_{\vec q}
R'(\omega) \Omega \rho_{\vec q} \rangle \right) \to i \vec q
\frac{\alpha}{S_0} \langle \delta \rho^*_{\vec q}(\omega) \rangle \; ,
\end{equation}
where the limit in the second part holds for $q\to0$ and $\omega\to0$,
and the coefficient $\alpha$ was defined 
in \gl{mi24}.  Therefore, in the hydrodynamic limit, if a density
gradient exists, it causes the force field: 
\begin{equation}\label{ltt2}
\langle \delta \vec F_{s}(\vec r,\omega) \rangle =
\frac{\alpha}{S_0}\; \nabla \; 
\langle \delta \rho(\vec r,\omega) \rangle \;.
\end{equation}

A sphere among the Brownian particles experiences this force density,
and if it moves with velocity $\vec v$, the density 
fluctuations in its vicinity are described by
\gls{mi4}{mi14}. Following, the macroscopic approach to long time tails,
and inserting these expressions (with obvious definition of the matrix
field $\alpha(\vec r,\omega)$) into \gl{ltt2} gives the force field around
the sphere. 
The sphere feels the interactions on its surface and thus the total force on
it is obtained by averaging over the surface: 
\begin{equation}\label{ltt3}
\vec F_{s,\omega} = \frac{n \alpha}{S_0} \vec v_\omega\; \left. \int
\frac{d^2f_r}{4\pi} ( \uvec v_\omega 
\cdot \nabla_r ) \int d^3s \uvec v_\omega \cdot \alpha(\vec s,\omega)
\cdot \nabla_r \; 
\Phi(\vec r-\vec s,\omega) \right|_{r=\sigma}\; ,
\end{equation}
where $\Phi(\vec r,\omega) = \frac{1}{4\pi Dr} e^{-\kappa r}$ with
$\kappa^2=-i\omega/D$ follows from \gl{mi13}.  In general this result
can not be simplified and e.g.\ the $\omega=0$ value, which would be
connected to the tracer diffusion coefficient, cannot be found from
our hydrodynamic consideration alone.  Linear response theory enables
one to identify the tracer force autocorrelation function from
\gl{ltt3}: $\vec F_{s,\omega} = -(1/3) \langle \vec F_s(\omega) \cdot
\vec F_s \rangle \; \vec v_{\omega}$.  In the dilute limit, it shows a
small frequency anomaly of order $\kappa^3$ and expecting this result
at finite densities also, we expand the fundamental solution $\Phi$ of
the diffusion equation up to this order. Abbreviating the
uninteresting terms this leads to:
\begin{equation}\label{ltt4}
\langle \vec F_s(\omega) \cdot \vec F_s \rangle = c + c' i \omega +
 \frac{n\alpha^2}{4\pi DS_0\sigma^3}
 \left[\frac{-i\omega\sigma^2}{D}\right]^{3/2}\; + {\cal
 O}(\omega^2)\; .
\end{equation}
Importantly, the linear term in $\kappa$ vanishes (it would indicate
$\langle \vec F_s(t\to\infty) \cdot \vec F_s \rangle \propto
t^{-3/2}$), and the leading $\omega^{3/2}$ anomaly corresponds to the
final power law decay $\langle \vec F_s(t\to\infty) \cdot \vec F_s
\rangle \sim 3\pi n\alpha^2/[16 S_0 (\pi Dt)^{5/2}]$. As was
expected from the spatial long--ranged pattern around the tracer there
exists a temporal long time tail whose amplitude is closely connected
to the latter.  Interestingly, standard mode coupling theory gives
different results \cite{Pusey85,Hess83,Goetze91b,Fuchs98} and requires
additional considerations \cite{kroy}. Our result implies that a
frequency dependent vertex coupling density modes would be required in
order to obtain \gl{ltt4}, which agrees with the known low density
results \cite{Ackerson82,Cichocki92} upon accounting for the tracer
diffusion by the replacement $D\to D_0+D_s$ and the identification
$\sigma=a+a_s$.  The tracer mean squared displacement is connected to
the force correlation function via the equation of motion,
$\partial_t^2 \langle \Delta r^2(t) \rangle = -2 D_s^2 \langle \vec
F_s(t) \cdot \vec F_s \rangle$, and exhibits a power law approach
$\propto -\alpha^2 t^{-1/2}$ to the long time diffusion.

\section{Conclusions and outlook}

We have presented the first statistical mechanics derivation of a
hydrodynamic boundary condition for the diffusion equation, including
the definition of the macroscopic boundary position from molecular
parameters. This has proven surprisingly difficult, because a
non--analytic (non-Markovian) structure in the resulting relaxation
kernel needed to be identified. Quoting exact low density results and
performing a mode coupling approximation, this structure could be
established. The $1/\sqrt\omega$ anomaly of diffusion close to a wall
is connected to the penetration length $\sqrt{D/\omega}$ which arises
generically in these situations \cite{Landau59}. Thus, we believe the
appearence of non--Markovian relaxation kernels is inherent to the
statistical mechanics derivation of boundary conditions. Gratifyingly,
within the mode coupling approximation it arises from a non--linear
coupling of the hydrodynamic modes themselves, and thus, as required
for a macroscopic hydrodynamic concept, does not depend on molecular
details.

As an application of the derived boundary condition, we calculated the
long range density pattern around, the resulting force on, and the
mean squared displacement of a tracer sphere moving among the bath
particles.  Extending results for dilute sytems, a generic power law
approach (long time tail) to the ultimate diffusion was found in the
motion of the tracer, whose amplitude ($\propto \alpha^2$) we
determined for arbitrary tracer size.  Interestingly, $\alpha$ may
vanish for special interaction parameters, in which case the tracer
might be called ``invisible''.

Clearly our calculation only presents a first step to determining
boundary conditions for hydrodynamic equations. Thus when considering
dispersed particles, their hydrodynamic interactions mediated via the
solvent should be included. As these are long--ranged, interesting
effects may appear. Also the description of the surface should be
improved by going beyond the single tracer calculation.  Structured or
rough surfaces could be modeled, as well as fluctuating ones in order
to adress the dynamics close to e.g.\ membranes. As the Smoluchowski
equation is the basis for the dynamics of polymers and general
macromolecules, the diffusion of complex molecules close to surfaces
also could be analyzed following the presented approach.  We hope to
address some of these points in the future.

\ack We want to thank M.E. Cates for his support and many helpful
discussions and his critical reading of the manuscript.  We also thank
L. Bocquet and J.-L. Barrat for valuable discussions.  M.F.\ was
supported by the Deutsche Forschungsgemeinschaft under grant Fu~309/3
and through the SFB 563, and K.K. by a European Community Marie Curie
Fellowship.

\appendix
\section{Solution of the diffusion equation around a sphere}

For convenience of comparison with the results obtained in the main
text, this abstract summarizes some hydrodynamic expressions. To
linear order in the perturbing tracer velocity $v$, it is
straightforward to solve the macroscopic \gls{e1}{e2} around a sphere
with radius $\sigma$ that oscillates with velocity amplitude $\vec
v_\omega$. With $\kappa^2 = -i\omega/D$ the square of the inverse skin
depth, one obtains
\begin{equation}\label{a2}
n_\omega(\vec r) = \vec v_\omega\cdot  \uvec r
\frac{n\; \sigma^3\,/D}{2+2\kappa \sigma + (\kappa \sigma)^2}
\frac{1+\kappa r}{r^2}e^{-\kappa(r-\sigma)}\;.
\end{equation}
It simplifies to the far--field expression
\begin{equation}\label{a3}
n_\omega(\vec r) \sim \vec v_\omega\cdot \uvec r \,
\frac{ n\; \sigma^3}{2D}
\frac{1+\kappa r}{r^2}e^{-\kappa r}\;, \qquad
n_{\vec q, \omega} \sim \vec v_\omega\cdot \vec q \, \frac{2\pi i\,
n\, \sigma^3}{Dq^2-i\omega} \; ,
\end{equation}
for a small sphere, $\sigma\to0$. Close to the sphere, it takes the
near--field expression
\begin{equation}\label{a4}
n_\omega(x) \sim  \frac nD \; {\rm v}_\omega^x
 \,\frac{ e^{-\kappa \bar x}}{\kappa} \;, \qquad
n_{q_x,\omega} \sim n\;{\rm v}_\omega^x
\;\frac{1+\sqrt{-iDq_x^2/\omega}}{Dq_x^2-i\omega} \; ,
\end{equation}
because the sphere degenerates to a plane upon taking the limit
$\sigma\to\infty$; here $\bar x$ gives the distance ($\vec r = \sigma
\uvec x +\bar{\vec r}$) and the Fourier transformation in \gl{a4} is
one--dimensional.

\section{Mobile tracer calculation}
To simplify the presentation in the main text and to make direct
contact with the hydrodynamic calculation, we have worked with a
non--fluctuating, macroscopic tracer throughout.  This appendix
extends the calculations of the main text to a finite tracer mobility
$D_s$, and ckecks that they are indeed recovered by taking the limit
$D_s\to0$.

Concerning the tracer velocity in  \gl{mi3_2};  
with \gls{mi5}{mi6}, and working in linear
response approximation, it becomes
\begin{equation}\label{eq:mi3_3}
\avg{\partial_t\vec r_s}^{\rm (ne)} = \avg{\vec v} +\avg{ D_s\vec
F_s}^{\rm (ne)} = \vec v -  D_s\!\int_{-\infty}^t\!\!\!\!\! d\tau\;
\vec v(\tau) 
\cdot \langle \vec F_s\, e^{\Omega (t-\tau)} \vec
F_s \rangle \;.
\end{equation}
The deviation from the hydrodynamic velocity $\vec v$ is explicitly of
the order $D_s$ and thus vanishes as required for $D_s=0$.

At finite $D_s$, 
the resolvents $R_{\vec q}$ and $R$ in Section~\ref{sec:tss} differ.
 Further, the
tagged particle density now being a dynamic flucutating varible, it
has to be considered as a separate slow mode in the projection in
\gls{mi9}{mi10}:
\begin{equation}
\label{eq:p}
P = \gva{\rho^s_{\vec q}}{}{\rho^{s*}_{\vec q}}
+\gva{\rho_{\vec q}}{\frac1{NS_{\vec q}}}{\rho^*_{\vec q}} -
\gva{\rho^s_{\vec q}}{\frac{n\, c^s_{\vec q}}{N}}{\rho^*_{\vec q}} -
\gva{\rho_{\vec q}}{\frac{n\, c^s_{\vec q}}{N}}{\rho^{s*}_{\vec q}} \;.
\end{equation}
Here, $c^s_{\vec q}=h^s_{\vec q}/S_{\vec q}$ is the (tagged) direct
correlation function. It is straightforward to check that $PP=P$,
$P\ket{\rho_{\vec q}}=\ket{\rho_{\vec q}}$, and $P\ket{\rho_{\vec
q}^s}=\ket{\rho_{\vec q}^s}$ up to corrections that are smaller by $(n
h^s_{\vec q})^2\!/(NS_{\vec q})$ relative to the leading order. The
susceptibility of \gl{mi12} is rewritten as
\begin{equation}\label{eq:suscept_latz}
n\chi_{\vec q}(\omega) = -
\left[ \avg{\vec F_{\vec q}^{s*} \rho_{\vec q}} 
-\avg{\vec F_{\vec q}^{s*} R_{\vec q}'\Omega_{\vec q}\rho_{\vec q}} \right]
\avg{\rho_{\vec q}^* R_{\vec q}\,\rho_{\vec q}}/(NS_{\vec q})  \;. 
\end{equation}
Hence, exactly the same decomposition of the susceptibility into an
instantaneous and a retarded contribution has been achieved as in
\gl{mi12} of the main text because of the negligible feedback of the
tracer onto the bulk \cite{Schofield92}.  To make contact with the
expressions in the main text, a factorization approximation is
required because of the tracer motion
\begin{equation}\label{eq:decoupling}
\avg{\rho_{\vec q}^* R_{\vec q}\,\rho_{\vec q}}=
\avg{\rho_{\vec q}^*\rho_{\vec q}^{s}R\rho_{\vec q}^{s*}\rho_{\vec q}} \approx
NS_{\vec q} \Phi_{\vec q}^{\phantom s}(\omega)\Phi_{\vec
q}^s(\omega)\; ,
\end{equation}
where $\Phi^s_{\vec q}(\omega) \equiv \avg{\rho_{\vec
q}^{s*}R(\omega)\rho_{\vec q}^s}$. The memory function $\avg{\vec
F_{\vec q}^{s*}R_{\vec q}'\Omega_{\vec q}\rho_{\vec q}}$ also acquires
a contribution from the tracer diffusion. To relevant lowest order in
$q$ it becomes
\begin{equation}\label{eq:gk}
 i D_0\avg{\vec F_{\vec q}^{s*}R_{\vec q}' \scal qF_{\vec q}} -
 iD_s\avg{\vec F_{\vec q}^{s*}R_{\vec q}'\scal qF_s\rho_{\vec q}}\;.
\end{equation}
The limit $D_s\to0$ recovers the results in the main text, where
$D_s=0$ from the outset. 

\vspace{\baselineskip}

\hrule

\vspace{\baselineskip}


\begin{thebibliography}{10}

\bibitem{hansen}
Hansen J P and McDonald I R
\newblock  1986 {\em Theory of Simple Liquids}
\newblock (Academic Press, London)

\bibitem{Onsager31}
Onsager L
\newblock  1931 {\em Phys. Rev.} {\bf 37} 405; {\bf 38} 2265

\bibitem{Kadanoff63}
Kadanoff L P and Martin P C
\newblock  1963 {\em Ann. Phys.} {\bf 24} 419

\bibitem{Landau59}
Landau L D and Lifshitz E M
\newblock  1959 {\em Fluid Mechanics}
\newblock (Pergamon Press, London)

\bibitem{Cercignani}
Cercignani C
\newblock  2000 {\em Rarefied Gas Dynamics: From Basic Concepts to Actual
  Calculations}
\newblock (Cambridge University Press, Cambridge)

\bibitem{Wolynes76}
Wolynes P G
\newblock  1976 {\em Phys. Rev. A} {\bf 13} 1235

\bibitem{Bocquet94}
Bocquet L and Barrat J.-L
\newblock  1994 {\em Phys. Rev. E} {\bf 49} 3079

\bibitem{Pusey85}
Pusey P N and Tough R. J A
\newblock  1985 {\em Dynamic Light Scattering: Application of Photon
  Correlation Spectroscopy} ed{} Pecora R
\newblock (Plenum, New York) p~85

\bibitem{dhont}
Dhont J. K G
\newblock  1996 {\em An introduction to dynamics of colloids}
\newblock (Elsevier Science, Amsterdam)

\bibitem{Goetze89b}
G{\"o}tze W and Latz A
\newblock  1989 {\em J. Phys.: Condens. Matter} {\bf 1} 4169

\bibitem{Dieterich79}
Dieterich W and Peschel I
\newblock  1979 {\em Physica} {\bf 95A} 225

\bibitem{Shapiro90}
Shapiro M, Brenner H  and Guell D C
\newblock  1990 {\em J. Colloid Interface Sci.} {\bf 136} 552

\bibitem{Hess83}
Hess W and Klein R
\newblock  1983 {\em Adv. Physics} {\bf 32} 173

\bibitem{Goetze91b}
G{\"o}tze W
\newblock  1991 {\em Liquids, Freezing and Glass Transition} eds{} Hansen J.-P,
  Levesque D  and Zinn-Justin J
\newblock (North-Holland, Amsterdam) p 287

\bibitem{keyes-masters85}
Keyes T and Masters A J
\newblock  1985 {\em Adv. Chem. Phys.} {\bf 58} 1


\bibitem{Schofield92}
Schofield J and Oppenheim I
\newblock  1992 {\em Physica} {\bf A 187} 210

\bibitem{Machta82}
Machta J and Oppenheim I
\newblock  1982 {\em Physica} {\bf A 112} 361

\bibitem{Ackerson82}
Ackerson B J and Fleishmann L
\newblock  1982 {\em J. Chem. Phys.} {\bf 76} 2675

\bibitem{Cichocki92}
Cichocki B and Felderhof B U
\newblock  1992 {\em Langmuir} {\bf 8} 2889

\bibitem{Zwanzig70}
Zwanzig R and Bixon M
\newblock  1970 {\em Phys. Rev. A} {\bf 2} 2005

\bibitem{Hagen97}
Hagen M H J, Pagonabarraga I, Lowe C P  and Frenkel D
\newblock  1997 {\em Phys. Rev. Lett.} {\bf 78} 3785

\bibitem{Fuchs98}
Fuchs M, G{\"o}tze W and Mayr, M R  
\newblock  1998 {\em Phys. Rev. E} {\bf 58} 3384

\bibitem{kroy} Kroy K unpublished (2002)

\end{thebibliography}

\end{document}